\documentstyle[aps,multicol,pre,epsfig]{revtex}

\input{psfig}

\begin{document}

\begin{title}
{\bf A Discrete Nonlinear Model with Substrate Feedback}
\end{title}

\author{P. G. Kevrekidis$^{1,2}$, B.A. Malomed$^{2,3}$ and A. R. Bishop$^{2}$}
\address{
$^1$ Department of Mathematics and Statistics, University of
Massachusetts, Amherst, MA 01003-4515, USA \\
$^2$ Center for NonLinear Studies and Theoretical Division, MS B258,
Los Alamos National Laboratory, Los Alamos, NM 87545, USA \\
$^3$ Department of Interdisciplinary Studies, Faculty of Engineering, Tel
Aviv University, Tel Aviv 69978, Israel
}
\date{\today}
\maketitle

\begin{abstract}
We consider a prototypical model in which a nonlinear field
(continuum or discrete) evolves on a flexible substrate which feeds 
back to the evolution of the main field.
We identify the underlying physics and potential applications of
such a model and examine its simplest one-dimensional Hamiltonian form,
which turns out to be a modified Frenkel-Kontorova model
coupled to an extra linear equation. We
find static kink solutions and study their stability, and then examine
moving kinks (the continuum limit of the model is studied too). 
We observe how the substrate effectively renormalizes
properties of the kinks. In particular, a
nontrivial finding is that branches of stable and unstable
kink solutions may be extended beyond a critical point at which
an effective intersite coupling vanishes; passing this critical
point does not destabilize the kink.
Kink-antikink
collisions are also studied, demonstrating alternation between
merger and transmission cases.
\end{abstract}

\begin{multicols}{2}

\section{Introduction}

Soft condensed matter systems, such as vesicles, microtubules and membranes,
have recently attracted a lot of attention in both biological and industrial
applications \cite{av1,av2,gaid1,gaid2}. More generally, physical systems at
the nanoscale including nanotubes and electronic and photonic waveguide
structures \cite{gaid3,gaid4} have nontrivial geometry and are influenced by
substrate effects. These wide classes of problems, many of which are
inherently nonlinear, raise the question of the interplay between
nonlinearity and an adaptive substrate, including, in particular, a
possibility of developing curvature in the substrate due to feedback forces
from the nonlinear system. Some of these systems such as, for instance, the
DNA double strand \cite{GaiReiss,PB}, are also inherently discrete.

In this situation, the derivation of evolutionary equations for intrinsic
fields in a nonlinear system interfacing with a flexible substrate should
take into regard the local dynamics and the feedback of the substrate. A
context where these phenomena can easily manifest themselves is elasticity.
Usually, models of particles connected by springs, such as the
Frenkel-Kontorova (FK) model \cite{FK}, assume that the springs are strictly
linear. However, in reality the spring constant may depend on the spring's
stretch. In this case, we are meaning not a straightforward generalization
of the FK model that includes the spring's anharmonicity, but a different
system, in which the spring constant obeys its own dynamical equation, see
below.

There is an increasing body of literature dealing with the interplay of
nonlinearity, discreteness, and a substrate subject to curvature. Usually,
even if the substrate is curved, its geometry is assumed to be {\it fixed},
see, e.g., Refs. \cite{curv}. However, for many applications, ranging from
condensed matter to optics to biophysics, it is quite relevant to examine a
prototypical model that admits a possibility of a variable substrate, which
is affected by the primary field(s) and feeds back into its (their) dynamics.

In this work, we introduce a simple one-dimensional model which captures
some of the physically essential features of the interplay of a nonlinear
system with a flexible substrate. A natural setting for the model is a
lattice with on-site nonlinearity, or its continuum counterpart (see below),
hence the first ingredient of the system is the FK model (alias the discrete
sine-Gordon equation \cite{PK}): 
\begin{equation}
\ddot{u}_{n}=C\Delta _{2}u_{n}-\sin u_{n},  \label{beq1}
\end{equation}
where $C$ is the coupling constant (the spring constant, if we consider
adjacent sites as being coupled by springs), $\Delta
_{2}u_{n}=(u_{n+1}+u_{n-1}-2u_{n})$ is the discrete Laplacian, while the
last term in Eq. (\ref{beq1}) is the on-site nonlinearity. The overdot
denotes the temporal derivative, while $n$ indexes the lattice sites. The FK
model has a plethora of physical
realizations, the simplest one being the Scott's model, i.e., a chain of
pendula suspended on a torque-elastic thread \cite{Scott}.

\begin{figure}[tbp]
\centerline{\psfig{file=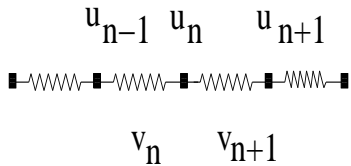,height=2in,width=2in,angle=270}}
\caption{Sketch of the model. $v_n$ is the deviation of the $n$-th spring
constant from $C_0$ due to the discrete gradient of the displacement $u_n$.}
\label{cfig1}
\end{figure}

The springs can be made nonlinear by assuming that the constant $C$ in Eq. 
(\ref{beq1}) is replaced by a {\em site-dependent} one, $C_{n}=C_{0}+v_{n}$,
where $C_{0}$ is the constant mean value, while $v_{n}$ is the variation of
the spring constant due to the variation of the displacements $u_{n}$ and 
$u_{n-1}$ at the sites that the spring couples. The simplest nontrivial
possibility, provided that we aim to produce a Hamiltonian system, is to
assume that $v_{n}$ responds to a change in displacements according to the
following equation: 
\begin{equation}
\ddot{v}_{n}=-\left[ \alpha v_{n}-k\left( u_{n}-u_{n-1}\right) \right] ,
\label{beq2}
\end{equation}
where $\alpha $ is an intrinsic stiffness of the spring (see below), and $k$
accounts for its susceptibility to the stretch. We use the convention (see
also a sketch of the system in Fig. \ref{cfig1}) that the $n$-th spring
connects the sites $n-1$ and $n$. To maintain the Hamiltonian character and
self-consistency of the model, the equation for the field $u_{n}$ should be
modified [cf. Eq. (\ref{beq1})] to read: 
\begin{equation}
\ddot{u}_{n}=C_{0}\Delta _{2}u_{n}-\sin u_{n}+k(v_{n}-v_{n+1}).  \label{beq3}
\end{equation}
In fact, the extra terms added to Eq. (\ref{beq3}) can be easily understood
in terms of the mechanical model: a difference in the value of the string
constant produces a difference in the elastic forces even if there is no
change in the length of the springs. Equations (\ref{beq2}) and (\ref{beq3})
are derived from the Hamiltonian 
\[
H=\sum_{n}\left[ \frac{1}{2}\left( \dot{u}_{n}^{2}+\dot{v}_{n}^{2}\right) +
\frac{\alpha }{2}v_{n}^{2}+\left( 1-\cos (u_{n})\right) \right. 
\]
\[
\left. +\frac{C_{0}}{2}(u_{n}-u_{n-1})^{2}-kv_{n}\left( u_{n}-u_{n-1}\right)
\right] . 
\]
We note that a somewhat similar (but still significantly different)
two-component model was proposed to describe the coupling between protons
and heavy ions in Ref. \cite{PPF} (see also references therein). A
single-component model with nonlinear springs was examined (mainly in terms
of its thermodynamic properties) in Ref. \cite{Kerr}.

While this work is focused on the simplest model based on Eqs. (\ref{beq2})
and (\ref{beq3}), which includes only the linear coupling between the fields 
$u_{n}$ and $v_{n}$, we note that a relevant generalization may involve
quadratic couplings: 
\begin{eqnarray}
\ddot{u}_{n} =C_{0}\Delta _{2}u_{n}-\sin (u_{n}) + k(v_{n}-v_{n+1}) 
\nonumber \\
+rv_{n}(u_{n-1}-u_{n})+rv_{n+1}(u_{n+1}-u_{n})\,,  \label{beq5} \\
\ddot{v}_{n} =-\left[ \alpha v_{n}-k\left( u_{n}-u_{n-1}\right) -\frac{r} {2}
\left( u_{n}-u_{n-1}\right) ^{2}\right].  \label{beq6}
\end{eqnarray}
The origin of the quadratic terms is evident: they stem from second-order
effects combining the change in the string length and elasticity.

The continuum limit (CL) of Eqs. (\ref{beq5}) and (\ref{beq6}) is 
\begin{eqnarray}
u_{tt} &=&u_{xx}-\sin u-kv_{x}+r(vu_{x})_{x}\,,  \label{beq7} \\
v_{tt} &=&-\alpha v+ku_{x}+\frac{r}{2}u_{x}^{2},  \label{beq8}
\end{eqnarray}
where the constants $k$ and $r$ were appropriately rescaled, and the
subscripts stand for partial derivatives. Setting $r=0$ in Eqs. (\ref{beq7})
and (\ref{beq8}) yields the CL of Eqs. (\ref{beq2}) and (\ref{beq3}). Note
that, even in the latter case, the Lorentz invariance of the CL equations is
broken (in comparison with the sine-Gordon equation, which is the CL of the
usual FK model). An exact solution for the static kink (the topological
soliton in the $u$--subsystem) can be easily found (for the case $r=0$; 
$\alpha $ is then rescaled to be 1), 
\begin{eqnarray}
u &=&4\tan ^{-1}\left[ \exp \left( \frac{x-x_{0}}{\sqrt{1-k^{2}}}\right)
\right] \,,  \label{beq9} \\
v &=&2\frac{k}{\sqrt{1-k^{2}}}{\rm sech}\left( \frac{x-x_{0}}{\sqrt{1-k^{2}}}
\right) \,,  \label{beq10}
\end{eqnarray}
but exact solutions for moving kinks are not available even in the CL limit.

Also worth noting is the CL for the linear spectrum of 
extended waves, which can
be obtained by substituting $u\sim A\exp (i(Kx-\Omega t))$ and $v\sim B\exp
(i(Kx-\Omega t))$ in the linearized version of Eqs. (\ref{beq7}) and (\ref
{beq8}). For $r=0$ and with the normalization $\alpha =1$, we obtain 
\[
\Omega =\pm \sqrt{1+\frac{1}{2}K\left( K\pm \sqrt{K^{2}+4k^{2}}\right) },
\]
which consists of two phonon bands, one with $\sqrt{1+|k|}<\Omega <\infty $,
and the other one with $\sqrt{1-|k|}<\Omega <1$. Notice the presence of a
finite gap between these bands, provided that $k\neq 0$. In the case $|k|=1$,
the lower branch of the linearized spectrum is acoustic, while the upper
one is always of the optical (alias, plasmon) type.

\begin{figure}[h]
\epsfxsize=5.5cm
\centerline{\epsffile{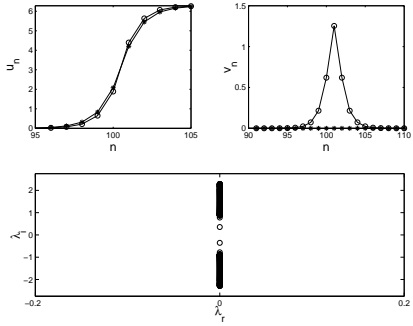}}
\caption{A stable static intersite-centered kink ($u_n$ and $v_n$ components
are shown by circles in the top left and top right plots, respectively) for 
$C_0=1$ and $k=0.5$; solid lines connecting circles are guides to the eye.
For comparison, the fields are also shown for the case $k=0$, which
corresponds to the usual FK model (stars connected by solid lines). The
bottom subplot shows the spectral plane $(\protect\lambda _r,\protect\lambda_
i)$ of linear stability eigenvalues $\protect\lambda$ found from the
linearization of Eqs. (\ref{beq2}) and (\ref{beq3}) around the kink (for 
$k=0.5$; the subscripts $i$ and $r$ refer to the real and imaginary parts of
the eigenvalues). The absence of eigenvalues with nonzero real part implies
the stability of the kink. Here, $\protect\alpha=1$.}
\label{cfig2}
\end{figure}

Returning to the discrete case, there is no explicit solution for the static
case; however, after the substitution of $v_{n}=\left( k/\alpha \right)
\left( u_{n}-u_{n-1}\right) $, which follows from Eq. (\ref{beq3}), the
steady-state equation for $u_{n}$ becomes 
\begin{equation}
(C_{0}-k^{2}/\alpha )\Delta _{2}u_{n}=\sin u_{n},  \label{static}
\end{equation}
and hence $k$ effectively renormalizes the lattice coupling. In fact that is
what can be observed from Eqs. (\ref{beq9}) and (\ref{beq10}), as well as
from Fig. \ref{cfig2}: the effect of $k$ is to decrease the width of the
kink (by a factor approximately equal to $|k|$, for $k$ small). As is well
known \cite{PK,KJ}, a narrower kink is less mobile. A numerical solution of
the static discrete equation (\ref{static}) yields, similar to the case of
the regular FK model \cite{PK,KJ}, two types of steady-state discrete kinks,
namely a stable inter-site centered one (an example for $C_{0}=2k=1$ is
shown in Fig. \ref{cfig2}) and an unstable site-centered one.

\begin{figure}[h]
\epsfxsize=5.5cm
\centerline{\epsffile{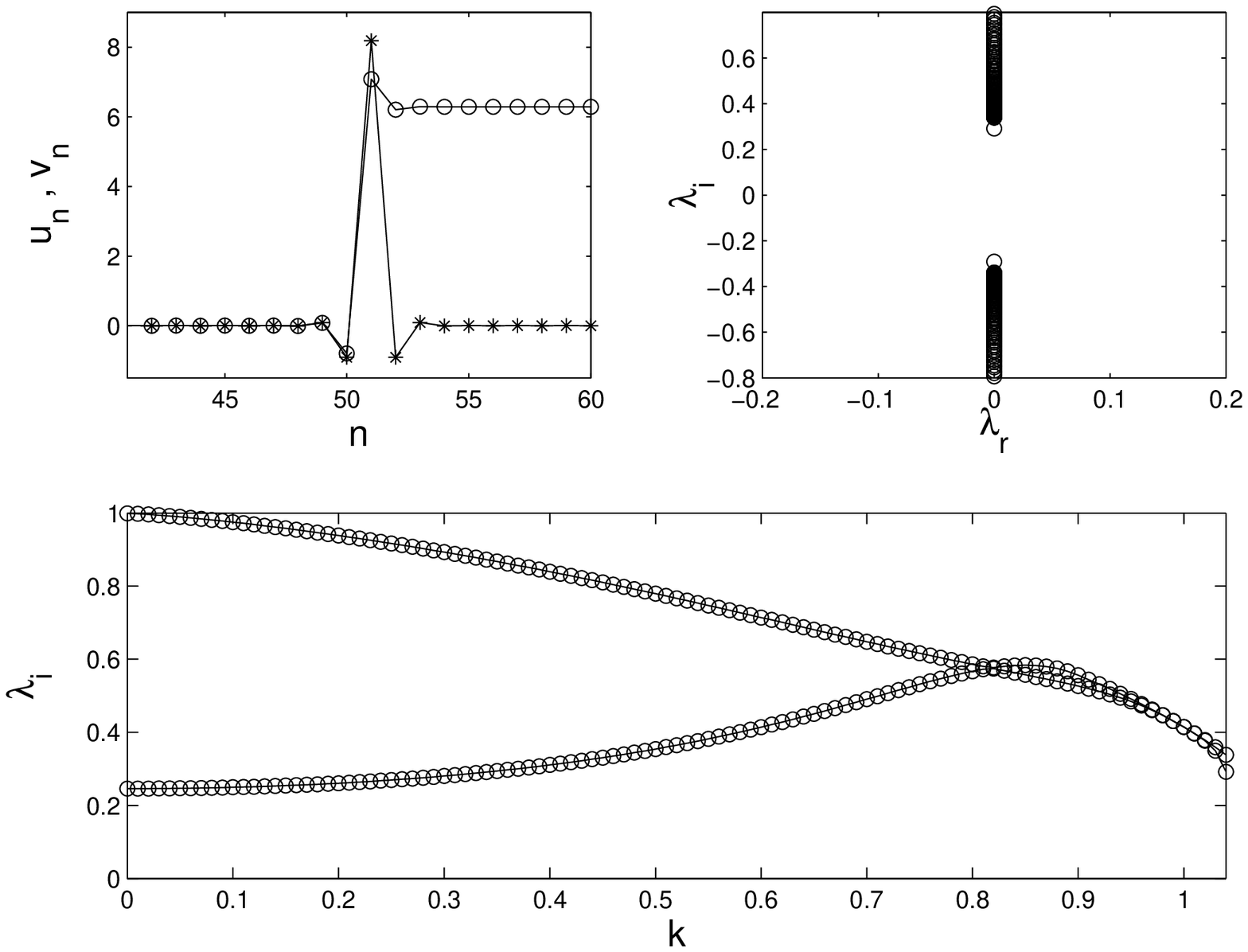}}
\epsfxsize=5.5cm
\centerline{\epsffile{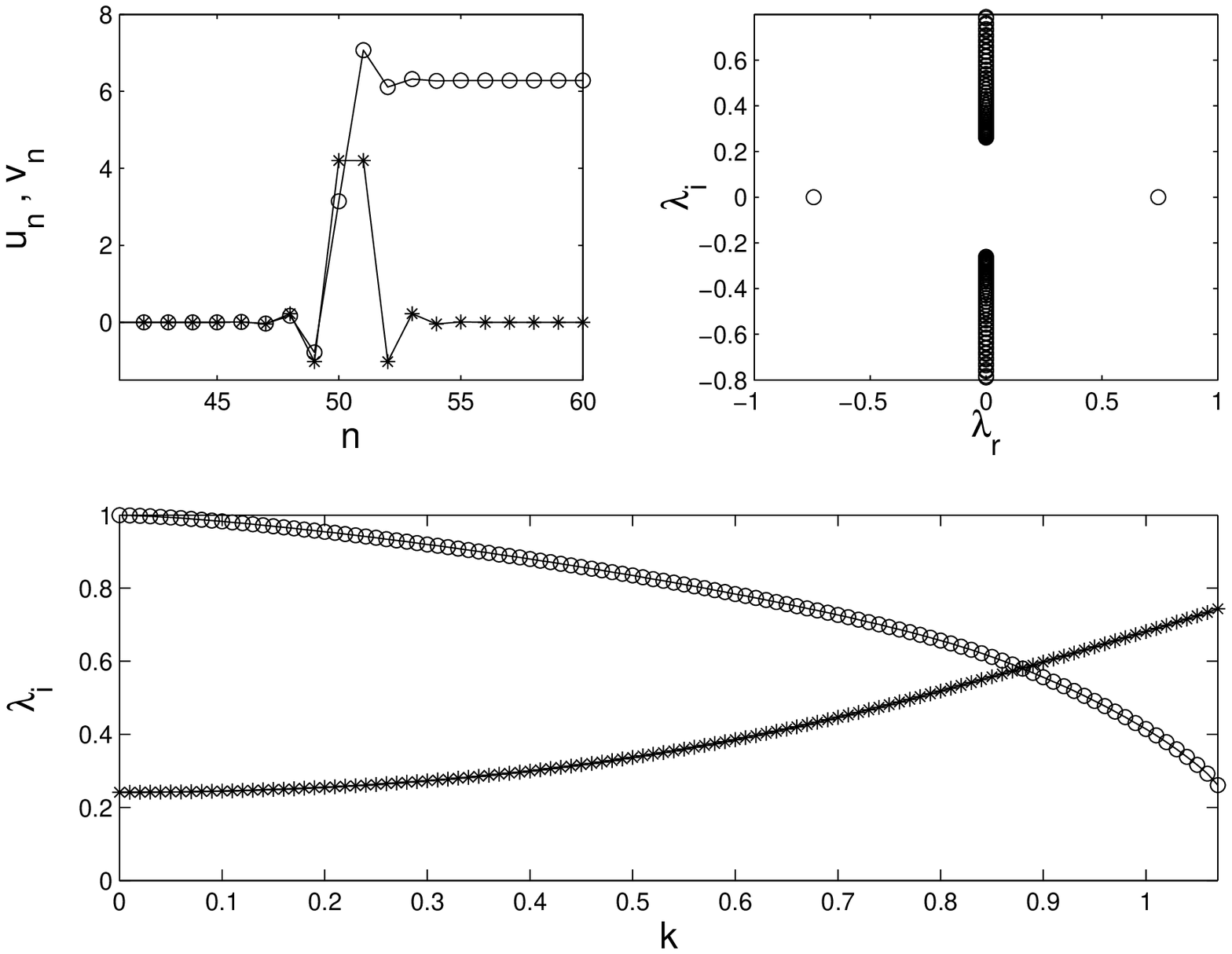}}
\caption{The top subplot shows the intersite-centered solution and its
stability (top left and right panels) for $|k|=1.04$ (just prior to the
branch termination). The bottom panel shows the two stable internal-mode
eigenvalues of the intersite-centered kink (circles connected by the solid
line) vs. $k$. The bottom subplots show the same for the unstable
site-centered solution (the two top panels are for $k=1.07$, i.e., also just
prior to the termination of the corresponding branch, and the stars in the
bottom subplot show the {unstable} eigenvalue). Note that the profiles of
both kinks are non-monotonic.}
\label{cfig3}
\end{figure}

We have examined the variation of the static kink solutions and of their
linear stability eigenvalues (for small perturbations) as a function of $|k|$.
Typical examples are shown in Fig. \ref{cfig3} for the stable and unstable
kinks for $C_{0}=\alpha =1$. A surprising finding in both cases is that the
solutions could be continued for some range {\it beyond} the point $|k|=1$,
which is effectively equivalent to the zero-coupling ({\it
anti-continuum}) limit, see Eq. (\ref{static}). The shape of the solutions
found for $\left| k\right| >1$ is non-monotonic (but they remain {\em stable}).
These solution branches terminate at $|k|\approx 1.04$ for the stable
solution, and at $|k|\approx 1.07$ for the unstable site-centered one.

\begin{figure}[h]
\epsfxsize=5.5cm
\centerline{\epsffile{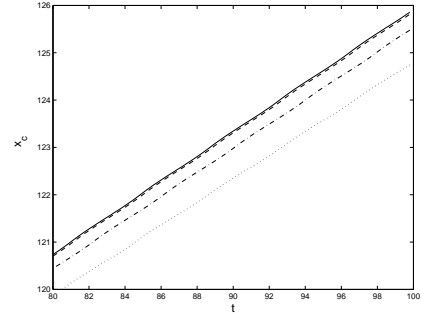}}
\caption{Trajectories of a continuum-limit kink, initially boosted to the
velocity $c=0.2$, obtained from the numerical integration of Eqs. (\ref{beq2})
and (\ref{beq3}) at different values of $k$. $k=0,\, k=-0.05,\, k=-0.15$,
and $k=-0.25$ correspond, respectively, to the solid, dashed, dash-dotted,
and dotted lines. The trajectories are shown only for late stage of the
time evolution.}
\label{cfig4}
\end{figure}

The above-mentioned expectation that the kink becomes less mobile as $|k|$
increases is verified dynamically by taking an initial kink boosted to the
speed $c=0.2$. Solving Eqs. (\ref{beq7}) and (\ref{beq8}) (with $r=0$) with
this initial condition for different values of $k$, we notice that, as is
shown in Fig. \ref{cfig4}, the kink travels farther for $k=0$ than it does
for $k\neq 0$, despite the fact that the initial velocity is the same in all
the cases. In other words, the effective velocity which the kink
demonstrates differs from that which was lent to it initially, and it also
depends on $|k|$.

\begin{figure}[h]
\epsfxsize=5.5cm
\centerline{\epsffile{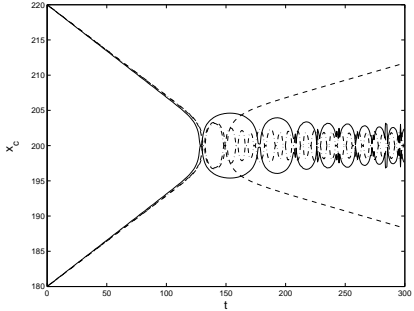}}
\caption{Kink-antikink collisions for $C_0=16/9$ ($h=0.75$)
and initial speeds 
$\pm 0.1$ are
shown in terms of trajectories of centers of the colliding kinks. 
The solid, dashed and dash-dotted lines correspond, respectively, to 
$k=0.0$, $k=0.178$ and $k=0.2$.}
\label{cfig5}
\end{figure}

Finally, we have also examined kink-antikink collisions in the system of
Eqs. (\ref{beq2}) and (\ref{beq3}). An example of this type is shown in Fig. 
\ref{cfig5}. In this case, $C_{0}=16/9$ ($h=0.75$), and the initial speeds
of the kink and antikink are $\pm 0.1$. The figure indicates the possibility
of existence of parameter windows for which the kinks eventually escape
each other's attraction (upon multiple bounces). Such windows
lie between intervals of merger (kink-antikink annihilation into a
breather). This feature is reminiscent of well-known
findings of Ref. \cite{Campbell} for kink-antikink collisions in continuum
(nonintegrable) models. However, unlike what was found in that work,
here the windows are in terms of the parameter $k$, observed at a fixed
value of the collision velocity. 

An example of such a window (a chacteristic case of which is shown 
in Fig. \ref{cfig5}) occurs
for $0.176 \leq k \leq 0.179$, even though in the 
FK model (e.g., for $k=0$, a case also shown in Fig. \ref{cfig5}), for
$v=0.1$,
only annihilation is possible. Notice that for this value ($0.1$) of the
initial speed (and within our resolution steps of $0.001$ in $k$),
no additional windows were identified. However, by considering 
different values of the initial speed (such as e.g., $v=0.125$), we 
have verified that typically multiple such windows can occur.
The dynamical behavior of the present model, which is essentially richer
than in its FK counterpart, is provided for
by the fact that the kink's internal
mode is affected by variations of the coupling to the substrate (see the
discussion above).

The case of $\alpha =0$ in Eq. (\ref{beq2}) has its own interest, as in this
case $v_{n}$ may be regarded as an acoustic phonon branch (the one without a
gap in its dispersion relation), which interacts with the optical (plasmon)
branch represented by the field $u_{n}$. Equations (\ref{beq2}) and (\ref
{beq3}) with $\alpha =0$ furnish an elementary example of this type of
interaction. In this case, the equations have {\it solely} uniform
zero-velocity solutions. Nontrivial solutions may only exist with a finite
speed $c$. In particular, the model's CL gives rise, in this case, to a
double ($4\pi $) kink (not shown here), which is produced by the equation 
$(1-c^{2})u_{\xi \xi }-\sin (u)+(k^{2}/c^{2})u_{\xi }=0$, where $\xi \equiv
x-ct$. However, this simplest version of the model with $\alpha =0$ is
flawed, as its Hamiltonian is unbounded from below, hence the model is not a
well-posed one. It can be amended to fix this problem, keeping the
optical-phonon and acoustic-phonon character of the fields $u_{n}$ and $v_{n}
$, but detailed consideration of this issue is beyond the scope of the
present work.

To conclude, in this work we have proposed a simple, prototypical model,
that may be developed to describe dynamics of kink-shaped excitations in
more complex systems. A noteworthy extension would be the
generalization of the model
to two dimensions, which may be very relevant for physical applications.
Furthermore, dissipative versions of the model can be of direct relevance to
chemical and biophysical applications. Detailed consideration of
higher-order feedback effects, collisions between kinks, and long-lived
breather-like states in this model would also be of interest.

This research was supported by the US Department of Energy under the
contract W-7405-ENG-36. 

\end{multicols}

\end{document}